\documentclass{moriond}

\def\be{\begin{equation}}
\def\ee{\end{equation}}
\def\bea{\begin{eqnarray}}
\def\eea{\end{eqnarray}}

\usepackage{slashed}

\newcommand{\gsim}{\lower.7ex\hbox{$\;\stackrel{\textstyle>}{\sim}\;$}}
\newcommand{\lsim}{\lower.7ex\hbox{$\;\stackrel{\textstyle<}{\sim}\;$}}
\newcommand{\cO}{{\mathcal O}}

\newcommand{\cC}{{\mathcal C}}
\newcommand{\cL}{{\mathcal L}}

\newcommand{\cF}{{\mathcal F}}

\newcommand{\cS}{{\mathcal S}}
\newcommand{\cT}{{\mathcal T}}
\newcommand{\cU}{{\mathcal U}}

\newcommand{\cZ}{{\mathcal Z}}
\newcommand{\SMrep}[3]{({\bf #1},\,{\bf #2},\,#3)}
\newcommand{\SMrepbar}[3]{({\bf \bar#1},\,{\bf #2},\,#3)}

\newcommand{\Photo}{}

\begin{document}
\vspace*{4cm}
\title{SEMILEPTONICS AT HIGH $p_T$}
\author{ D.A. FAROUGHY}
\address{Physik-Institut, Universit\"at Z\"urich, CH-8057 Z\"urich, Switzerland}

\maketitle\abstracts{
We consider the Drell-Yan processes $pp \to \ell \ell^{(\prime)}$ and $pp\to \ell \bar{\nu}$ at high-$p_T$ as probes of New Physics effects in semileptonic transitions. For this purpose, we describe the $2\to2$ scattering amplitudes in terms of general form-factors which we match to the Standard Model Effective Field Theory (SMEFT) at $\cO(1/\Lambda^{4})$ including corrections from dimension-$8$ operators, and to tree-level mediators with $\cO(\rm{TeV})$ masses arising from ultraviolet models. By using the latest LHC run-II data from monolepton and dilepton production channels, we derive constraints on the SMEFT and on leptoquark models, with the most general flavor structures. Our results are compiled into the {\tt Mathematica} package {\tt HighPT}, which provides a simple way to perform high-$p_T$ collider analyses of Drell-Yan production Beyond the Standard Model (BSM). This contribution is based on \cite{Allwicher:2022gkm,Allwicher:2022mcg}.}

%=====================================================
\section{Drell-Yan Amplitudes}
\label{subsec:amp_dec}
%=====================================================

\subsection{Amplitudes decomposition}
\begin{figure}[t!]
\begin{center}
    \includegraphics[width=.2\linewidth]{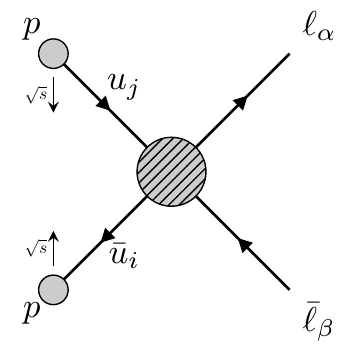}\hspace{1cm}
    \includegraphics[width=.2\linewidth]{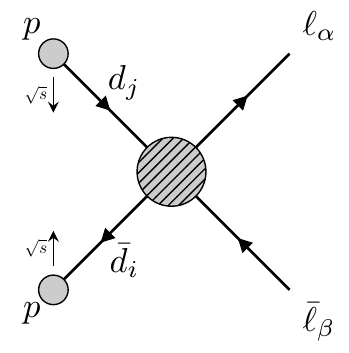}\hspace{1cm}
    \includegraphics[width=.2\linewidth]{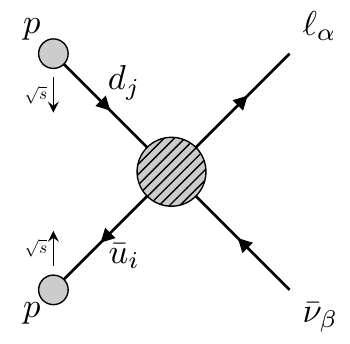}
    \caption{\sl\small Neutral and charged Drell-Yan production processes at proton-proton colliders.}
    \label{fig:DY_diags}
\end{center}
\end{figure}

First, we consider the scattering amplitude for the neutral Drell-Yan process $\bar{q}_i {q}_j \to \ell_\alpha^- \ell_\beta^+$ given by the first two diagrams in Fig.~\ref{fig:DY_diags}, with $q_i=\{u_i,d_i\}$, where quark and lepton flavor indices are denoted by Latin letters ($i,j=1,2,3$) and Greek letters ($\alpha,\beta=1,2,3$), respectively \footnote{For up-type quarks the the indices run as $i,j=1,2$ because of the negligible top-quark content of the proton at LHC energies.}. The most general decomposition of the four-point scattering amplitude that is Lorentz and gauge invariant is given by  
\begin{eqnarray}
\nonumber
\mathcal{A}(\bar{q}_i q_j \to {\ell}_{\alpha}^-\ell^+_{\beta})\, =\, \frac{1}{v^2}\,\sum_{XY} &\Big\lbrace&
\left(\bar \ell_\alpha\gamma^\mu   P_X \ell_\beta\right)\left(\bar q_i\gamma_\mu  P_Y q_j\right)\, [\mathcal{F}^{XY,\,qq}_{V}(\hat{s},\hat{t})]^{\alpha\beta}_{ij}\\\nonumber
&+& \left(\bar \ell_\alpha  P_X\ell_\beta\right)\left(\bar q_i  P_Y q_j\right)\, [\mathcal{F}^{XY,\,qq}_{S}(\hat{s},\hat{t})]^{\alpha\beta}_{ij}\\ \label{eq:dilep-amp}
&+& \left(\bar \ell_\alpha\sigma_{\mu\nu} P_X\ell_\beta\right)\left(\bar q_i \sigma^{\mu\nu}  P_Y q_j\right) \, [\mathcal{F}^{XY,\,qq}_T(\hat{s},\hat{t})]^{\alpha\beta}_{ij}\\\nonumber
&+&  \left(\bar \ell_\alpha\gamma_{\mu} P_X\ell_\beta\right)\left(\bar q_i \sigma^{\mu\nu}  P_Y q_j\right)\,\frac{ik_\nu}{v}  \, [\mathcal{F}^{XY,\,qq}_{D_q}(\hat{s},\hat{t})]^{\alpha\beta}_{ij} \\\nonumber
&+& \left(\bar \ell_\alpha\sigma^{\mu\nu}   P_X \ell_\beta\right) \left(\bar q_i \gamma_{\mu}   P_Y q_j\right)\,\frac{ik_\nu}{v} \, [\mathcal{F}^{XY,\,qq}_{D_\ell}(\hat{s},\hat{t})]^{\alpha\beta}_{ij}\ \Big\rbrace\,,\nonumber
\end{eqnarray}
where $X,Y\!\in\!\{L,R\}$ are the chiralities of the anti-lepton and anti-quark fields, $P_{R,L}=(1\pm\gamma^5)/2$ are the chirality projectors, $v=(\sqrt{2}G_F)^{-1/2}$ stands for the electroweak vacuum-expectation-value (vev) and fermion masses have been neglected. Here it is understood that $q$ ($\bar q$) and $\ell$ ($\bar\ell$) denote the Dirac spinors of the incoming quark (anti-quark) and outgoing anti-lepton (lepton) fields, respectively. The four-momentum of the dilepton system is defined by $k = p_q+p_{\bar{q}}$, and we take the Mandelstam variables to be $\hat s = k^2= (p_q+p_{\bar q})^2$, $\hat t = (p_{q}-p_{\ell^-})^2$ and $\hat u= (p_{q}-p_{\ell^+})^2=-\hat s-\hat t$ for massless external states. For each of the five components in eq.~(\ref{eq:dilep-amp}) we define the neutral current form-factor $\cF_{I}^{XY,\,qq}(\hat s,\hat t)$ where $I\in\{V,S,T,D_\ell,D_q\}$ labels the corresponding {\em vector}, {\em scalar}, {\em tensor}, {\em lepton-dipole} and {\em quark-dipole} Lorentz structures, respectively. These form-factors are dimensionless functions of the Mandelstam variables that describe the underlying local and non-local semileptonic interactions between fermions with fixed flavors and chiralities. An almost identical expression can be derived for the charged current Drell-Yan process $\bar{u}_i d_j \to {\ell}_{\alpha}^-\bar\nu_{\beta}$ given by the last diagram in Fig.~\ref{fig:DY_diags}, with the same five Lorentz structures as in the previous case, with form-factors denoted by $\smash{\cF^{XY,\,ud}_I(\hat s,\hat t)}$. The above equation is also valid for $X=R$ in the presence of a light right-handed neutrino field $\nu_R$ that is a singlet under the SM gauge group. 

%%%%%%%%%%%%%%%%%%%
%%%%%%%%%%%%%%%%%%%

\subsection{Form-factor parametrization}
The form-factors introduced above can be continued analytically to complex Mandelstam variables. We assume these functions to be analytic in each Mandelstam variable inside the radius $\Lambda^2$ except for a finite set of simple poles in the $\hat s$ and $\hat t$ and $\hat u$ complex planes. Each form-factor can therefore be decomposed into a ``regular" term and a ``pole" term,
\begin{equation}\label{eq:FF}
 \cF_{I}(\hat  s,\hat{t}) =\ \cF_{I,\, \rm Reg}(\hat s,\hat{t})\ +\ \cF_{I,\, \rm Poles}(\hat s,\hat{t})\,,    
\end{equation}
that encode possible local and non-local interactions, respectively. These admit the following parametrization: 
\begin{eqnarray}\label{eq:regular}
\cF_{I,\,\rm Reg}(\hat  s,\hat{t}) &=&\sum_{n,m=0}^\infty \cF_{I \,(n,m)}\,\left(\frac{\hat s}{v^2}\right)^{\!n}\left(\frac{\hat t}{v^2}\right)^{\!m}\\
\cF_{I,\,\rm Poles}(\hat  s,\hat{t}) &=& \sum_a\frac{v^2\, \cS_{\,I\,(a)} }{\hat{s}-\Omega_a}
\ +\  \sum_b\frac{v^2\, \cT_{\,I\,(b)} }{\hat{t}-\Omega_b}
\ -\  \sum_c\frac{v^2\, \cU_{\,I\,(c)} }{\hat{s} + \hat {t} +\Omega_c}\,.\label{eq:poles}
\end{eqnarray}
Since the regular term $\cF_{I,\, \rm Reg}$ is an analytic function,  equation (\ref{eq:regular}) corresponds to a power series expansion valid for $|\hat s|,|\hat t|\ll\Lambda$ where the (dimensionless) coefficients $\cF_{I\,(n,m)}$ encode the effects of unresolved degrees of freedom living at the scale $\Lambda$ beyond the characteristic energy of the scattering process. This power series is not the complete Effective Field Theory (EFT) expansion in $1/\Lambda$ since $\cF_{I \,(n,m)}$ are ``all-order" coefficients that receive contributions from an infinite tower of non-renormalizable operators. The pole term $\cF_{I,\, \rm Poles}$, on the other hand, is a non-analytic function with simple poles located at $\Omega_k=m_k^2-im_k\Gamma_k$ in each of the complex Mandelstam planes. This function describes the tree-level exchange of possible bosonic degrees of freedom in the $s$-channel, $t$-channel and $u$-channel, respectively, i.e. the  propagators of various particles $a,b,c$ with masses $m_{a,b,c}$ and widths $\Gamma_{a,b,c}$ that can be resolved at the energies taking place in the scattering. The pole residues $\mathcal{S}_{I\,(a)}$, $\mathcal{T}_{I\,(b)}$ and $\mathcal{U}_{I\, (c)}$ are dimensionless quantities parametrizing local three-point interactions. In full generality, the simple pole assumption for the singular structures of the form-factors still allows for these residues to be analytic functions in each variable, i.e. $\mathcal{S}_{I\,(a)}(\hat s)$, $\mathcal{T}_{I\,(b)}(\hat t)$ and $\mathcal{U}_{I\, (c)}(\hat u)$. However, their dependence on the Mandelstam variables can be completely removed by using the identity
\begin{equation}\label{eq:trick}
\frac{\cZ(\hat z)}{\hat z-\Omega}\ =\ \frac{\cZ(\Omega)}{\hat z-\Omega}\ +\ h(\hat z,\Omega)\,,   
\end{equation}
\noindent where $h(\hat z,\Omega)$ is an analytic function of $\hat z=\{\hat s,\hat t,\hat u\}$ that can be reabsorbed into the regular form-factor by redefining $\cF_{I,\, \rm Reg}$. This can be seen by power expanding the numerator on the left-hand side of eq.~(\ref{eq:trick}) and applying the partial fraction decomposition $\hat z^n/(\hat z-\Omega)=\Omega^n/(\hat z-\Omega)+\Omega^{n-1}\sum_k^{n-1}(\hat z/\Omega)^k$ on each of the resulting terms. Using this simple identity allows to reduce considerably the number of parameters in $\cF_{I,\, \rm Poles}$ to just two parameters per pole: the location $\Omega$ of the pole and its residue $\cZ$. When discussing the SMEFT, we will see that this turns out to be useful in characterizing dimension-8 operators, in particular effects giving rise to energy enhanced three-point interactions.

\section{Semileptonics Beyond the SM}

%=====================================================
\subsection{Form-factors in the SMEFT at $\cO(1/\Lambda^4)$}
%=====================================================
In the SMEFT the regular form-factor coefficients $\cF_{I\,(n,m)}$ in eq.~(\ref{eq:regular}) are given by an expansion series of the form
\begin{eqnarray}
\label{eq:Fnm_expansion}
\cF_{I\,(n,m)}  &=& \sum_{ d\,\ge\,2(n+m+3)}^{\infty}\, c_I^{(d)} \left(\frac{v}{\Lambda}\right)^{d-4}\,,
\end{eqnarray}
where $d\ge4$, and $c_I^{(d)}$ are linear combinations of $d$-dimensional Wilson coefficients. Notice that SMEFT operators at a fixed mass dimension give rise to a finite number of form-factor coefficients. For example, dimension-$6$ operators only contribute to the leading coefficient $\cF_{I\,(0,0)}$, whereas dimension-$8$ operators contribute to $\cF_{I\,(0,0)}$, $\cF_{I\,(1,0)}$ and $\cF_{I\,(0,1)}$, and so on. For Drell-Yan production at $\cO(1/\Lambda^4)$ the relevant operator classes are {\bf$\psi^2H^2D$}, {\bf$\psi^2HX$} and {\bf$\psi^4$} at dimension-$6$, and {\bf$\psi^2H^4D$}, {\bf$\psi^2H^2D^3$}, {\bf$\psi^4H^2$} and {\bf$\psi^4D^2$} at dimension-$8$, defined in Refs.~\cite{Grzadkowski:2010es,Murphy:2020rsh}. 

In order to match the form-factor coefficients to the Wilson coefficients at $\cO(1/\Lambda^4)$ it is sufficient to truncate the expansion series of $\cF_{I,\,\rm Reg}$ in eq.~(\ref{eq:regular}) at $n,m\le1$. The regular pieces of the scalar and tensor form-factors can be further truncated at $n=m=0$ because when squaring the amplitude the terms with $n+m=1$ generated at dimension-$8$ do not interfere with the SM poles and will only lead to higher order effects beyond $\cO(1/\Lambda^4)$. We can set the regular form-factors $\cF_{D_\ell,\,\rm Reg}$ and $\cF_{D_q,\,\rm Reg}$ to zero since in the SMEFT the dipoles only arise from non-local interactions involving SM gauge bosons. For the pole form-factors $\cF_{I,\,\rm Poles}$, we only need to consider the vector poles and dipoles arising from the $s$-channel exchange of the SM gauge bosons. Furthermore, the $s$-channel scalar pole $\cF_{S,\,\rm Poles}$ generated from the exchange of the SM Higgs boson is completely negligible because of the small fermion Yukawa couplings of the external states. 

The coefficients $\cF_{S\,(0,0)}$ and $\cF_{T\, (0,0)}$  map directly to the Wilson coefficients of scalar and tensor operators in class {\bf$\psi^4$}, respectively. The dipole residues $\cS_{D\,(a)}$ also match trivially to the $d=6$ SMEFT dipole operators in class {\bf$\psi^2XH$}. The regular coefficients and the pole residues of the vector form-factors, on the other hand, match to operators at dimension-$6$ and dimension-$8$. The leading coefficient $\cF_{V\,(0,0)}$ receives contributions from contact operators in the classes {\bf$\psi^4$} and {\bf$\psi^4H^2$} at dimension-$6$ and dimension-$8$, respectively, as well as from modified interactions between fermions and the SM gauge bosons from dimension-$8$ operators in class {\bf $\psi^2H^2D^3$}. The higher order coefficients $\cF_{V\,(1,0)}$ and $\cF_{V\,(0,1)}$ receive contributions from the dimension-$8$ operators in class {\bf$\psi^4D^2$}. The pole residues $\delta \cS_{(a)}$ receive dominant contributions from modified vertices from dimension-$6$ operators in class {\bf$\psi^2H^2D$} and from dimension-$8$ operators in class {\bf$\psi^2H^4D$} and {\bf$\psi^2H^2D^3$}. Schematically, the matching between Wilson coefficients and the form-factors takes the following form: 
\bea
\cF_{V\,(0,0)}  &=& \frac{v^2}{\Lambda^2}\,\cC^{\,(6)}_{\psi^4}\ +\ \frac{v^4}{\Lambda^4}\,\cC^{\,(8)}_{\psi^4H^2}\ +\ \frac{v^2 m_a^2}{\Lambda^4}\,\cC^{\,(8)}_{\psi^2H^2D^3}\ + \ \cdots \,, \label{eq:FV00}\\
\cF_{V\,(1,0)}  &=& \frac{v^4}{\Lambda^4}\,\cC^{\,(8)}_{\psi^4D^2}\ + \ \cdots \,, \label{eq:FV10}\\
\cF_{V\,(0,1)}  &=& \frac{v^4}{\Lambda^4}\,\cC^{\,(8)}_{\psi^4D^2}\ + \ \cdots \,, \label{eq:FV01}\\
\delta \cS_{(a)} &=& \frac{m_a^2}{\Lambda^2}\,\cC^{\,(6)}_{\psi^2H^2D}\,+\frac{v^2m_a^2}{\Lambda^4}\,\left(\left[\cC_{\psi^2H^2D}^{\,(6)}\right]^2+\cC_{\psi^2H^4D}^{\,(8)}\right)\  +\ \frac{m_a^4}{\Lambda^4}\,\cC_{\psi^2H^2D^3}^{\,(8)}\ + \ \cdots\,, \label{eq:delF}
\eea
where the squared term $[\cC_{\psi^2H^2D}^{\,(6)}]^2$ in eq.~(\ref{eq:delF}) corresponds to double vertex insertions of the corresponding dimension-6 operator. The dots indicate negligible contributions from dimension-$10$ operators. For the full matching see~\cite{Allwicher:2022gkm}. Notice that the operators in class {\bf$\psi^2H^2D^3$} contribute to $\cF_{V\,(0,0)}$ and $\delta \cS_{(a)}$. This can be understood by analyzing one of the operator in this class, e.g. $\cO_{q^2H^2D^3}^{(1)}=(\bar q_i\gamma^\mu D^\nu q_j)D_{(\mu} D_{\nu)}H^\dagger H$. After spontaneous symmetry breaking, this operator gives rise to a modified coupling between the $Z$ boson and quarks that is proportional to $(\hat s\, m_Zv/\Lambda^4) \, Z_\mu(\bar q_i\gamma^\mu q_j)$. This interaction enters neutral Drell-Yan production with an amplitude that scales as $\mathcal{A}(\bar{q}_i q_j \to \ell^-_\alpha \ell^+_\beta)\propto \hat s/(\hat s - m_Z^2)$. This amplitude can be brought to the form eq.~(\ref{eq:FV00}-\ref{eq:delF}) by using the partial fraction decomposition in eq.~(\ref{eq:trick}).

%%%%%%%%%%%%%%%%%%%%%%%%%%%%%%%%%%%%%%%%%%%%%%%%%%%%%%%%%%%%%%%%%%%%%%%%%%
\subsection{Concrete UV Mediators}
%%%%%%%%%%%%%%%%%%%%%%%%%%%%%%%%%%%%%%%%%%%%%%%%%%%%%%%%%%%%%%%%%%%%%%%%%%

%%%%%%%%%%%%%%%%
\begin{table}[b!]
  \begin{center}
  \caption{Possible bosonic mediators contributing at tree level to Drell-Yan production. In the last column we provide the interaction Lagrangian where $\epsilon\equiv i\tau_2$, $\psi^c\equiv i\gamma_2\gamma_0\bar\psi^T$ and $\widetilde H=i\tau_2 H^\ast$ is the conjugate Higgs doublet. The right-handed fermion fields are defined as $u\equiv u_R$, $d\equiv d_R$, $e\equiv \ell_R$ and $N \equiv \nu_R$, and the left-handed fermion fields as $q\equiv (V^\dagger u_L,d_L)^T$ and $l\equiv (\nu_L,\ell_L)^T$. }
  {\renewcommand{\arraystretch}{1.3}
    \begin{tabular}{|c|c|c|l|}%{|@{\hspace{1em}}c@{\hspace{1em}}|c@{\hspace{1em}}|c@{\hspace{1em}}|l@{\hspace{0.1em}}|}
    \hline
    \ &   SM rep.  & Spin  & \hspace{3cm} $\cL_{\rm int}$\\
\hline
$Z^\prime$ & \SMrep{1}{1}{0} & 1 & $\cL_{Z^\prime}=\sum_\psi\, [g^{\psi}_1]_{ab} \, \bar \psi_a \slashed Z^\prime \psi_b$\, , \  $\psi\in\{u,d,e,q,l\}$\\
$W^\prime$ & \SMrep{1}{3}{0} & 1 & $\cL_{W^\prime}=[g^{q}_3]_{ij} \, \bar q_i \slashed W^\prime q_j+[g^{l}_3]_{\alpha\beta} \, \bar l_\alpha \slashed W^\prime l_\beta$ \\
%
% & & & \hspace{2.7cm}$\cL_{\rm int} + {\rm h.c.}$ \\
$\widetilde Z$ & \SMrep{1}{1}{1} & 1 & $\cL_{\widetilde Z}=[\widetilde g^{q}_1]_{ij}\, \bar u_i\slashed {\widetilde Z}d_j+[\widetilde g^{\ell}_1]_{\alpha\beta}\, \bar e_\alpha\slashed {\widetilde Z} N_\beta$\\
$\Phi_{1,2}$ & \SMrep{1}{2}{1/2}& 0 &$\cL_{\Phi}=\displaystyle\sum_{a=1,2}\Big{\lbrace} [y_{u}^{(a)}]_{ij}\, \bar q_i u_j\widetilde\Phi_a+[y_{d}^{(a)}]_{ij}\, \bar q_i d_j\Phi_a+[y_{e}^{(a)}]_{\alpha\beta}\, \bar l_\alpha e_\beta\Phi_a\Big{\rbrace}+\mathrm{h.c.}$\\
\hline

$S_1$ & $\SMrepbar{3}{1}{1/3}$  & 0 & $\cL_{S_1}= [y_1^L]_{i\alpha}\, S_1 \bar  q^c_i\epsilon l_\alpha+[y_1^R]_{i\alpha}\,S_1\bar u^c_i e_\alpha + {\  [\bar y_1^R]_{i\alpha}\,S_1\bar d^c_i N_\alpha}+\mathrm{h.c.}$\\ 
$\widetilde S_1$ & $\SMrepbar{3}{1}{4/3}$ & 0 & $\cL_{\widetilde S_1}=[\widetilde y_1^R]_{i\alpha}\,  \widetilde S_1\bar d^c_i e_\alpha+\mathrm{h.c.}$\\ 
$U_1$ & $\SMrep{3}{1}{2/3}$ & 1 & $\cL_{U_1}=[x_1^L]_{i\alpha} \, \bar q_i \slashed  U_{\!1} l_\alpha + [x_1^R]_{i\alpha} \,\bar d_i \slashed U_{\!1} e_\alpha + { [\bar x_1^R]_{i\alpha} \,\bar u_i \slashed U_{\!1} N_\alpha}+\mathrm{h.c.}$\\ 
$\widetilde U_1$ & $$\SMrep{3}{1}{5/3}$$ & 1 & $\cL_{\widetilde U_1}= [\widetilde x_1^R]_{i\alpha} \,\bar u_i \slashed {\widetilde U}_{\!1} e_\alpha+\mathrm{h.c.}$\\ 
$R_2$ & \SMrep{3}{2}{7/6} & 0  & $\cL_{R_2}= -[y_2^L]_{i\alpha} \, \bar u_i R_2 \epsilon l_\alpha+[y_2^R]_{i\alpha} \, \bar q_i e_\alpha R_2+\mathrm{h.c.}$ \\ 
$\widetilde R_2$ & \SMrep{3}{2}{1/6} & 0 & $\cL_{\widetilde R_2}=- [\widetilde y_2^L]_{i\alpha} \, \bar d_i \widetilde R_2 \epsilon l_\alpha+ { [\widetilde y_2^R]_{i\alpha} \, \bar q_i N_\alpha \widetilde R_2} +\mathrm{h.c.}$ \\ 
$V_2$ & $\SMrepbar{3}{2}{5/6}$ & 1 & $\cL_{V_{\!2}}=[x_2^L]_{i\alpha} \,\bar d_i^c \slashed V_{\!\!2}\epsilon l_\alpha+[x_2^R]_{i\alpha}\, \bar q_i^c\epsilon \slashed V_{\!\!2} e_\alpha +\mathrm{h.c.}$\\ 
$\widetilde V_2$ & $\SMrepbar{3}{2}{-1/6}$  & 1 & $\cL_{\widetilde V_{\!2}}=[\widetilde x_2^L]_{i\alpha} \,\bar u_i^c \slashed {\widetilde V}_{\!\!2}\epsilon l_\alpha+ { [\widetilde x_2^R]_{i\alpha} \,\bar q_i^c \epsilon \slashed {\widetilde V}_{\!\!2} N_\alpha} +\mathrm{h.c.}$\\ 
$S_3$ & $\SMrepbar{3}{3}{1/3}$ & 0 & $\cL_{S_3}=[y_3^L]_{i\alpha} \, \bar q^c_i \epsilon S_3 l_{\alpha}+\mathrm{h.c.}$\\ 
$ U_3$ & $\SMrep{3}{3}{2/3}$ & 1 & $\cL_{U_3}=[x_3^L]_{i\alpha} \, \bar q_i \slashed U_{3} l_\alpha+\mathrm{h.c.}$ \\
\hline
\end{tabular}
}
\label{tab:mediators}
\end{center}
\end{table}
%%%%%%%%%%%%%%%%

We now discuss the effects of new bosonic states exchanged at tree level in Drell-Yan production. These states can be classified in terms of their spin and SM quantum numbers. The possible mediators are displayed in Table~\ref{tab:mediators}, where we also show the relevant interaction Lagrangians with generic couplings in the last column. For completeness, we also allow for three right-handed neutrinos, denoted as $N_\alpha\sim({\bf 1}, {\bf 1},0)$, with $\alpha=1,2,3$. Furthermore, we assume that the masses of these SM singlets are negligible compared to the collider energies and, if produced, they can escape the detector as missing energy. The possible mediators fall into two broad categories: (i) color-singlets exchanged in the $s$-channel, and (ii) color-triplets, i.e. {\it leptoquarks}  \cite{Buchmuller:1986zs,Dorsner:2016wpm}  exchanged in the $t$- and $u$-channels. If the masses of these states are at the $\cO(\mathrm{TeV})$ scale their propagators will contribute to the residues $\cS_{I\,(a)}$, $\cT_{I\,(b)}$, $\cU_{I\,(c)}$ of the pole form-factors in (\ref{eq:poles}). Leptoquarks can be further classified using fermion number \cite{Dorsner:2016wpm}, defined as $F\equiv3B+L$ where $B$ ($L$) stands for Baryon (Lepton) number. For Drell-Yan production, the leptoquarks with no fermion number $F=0$, such as $U_1$, $\widetilde U_1$, $R_2$, $\widetilde R_2$ and $U_3$, are exchanged in the $t$-channel, while the remaining leptoquarks $S_1$, $\widetilde S_1$, $V_2$, $\widetilde V_2$ and $S_3$ carrying fermion number $F=-2$ are exchanged in the $u$-channel. Notice that these last states can also couple to diquark bilinears of the form $\bar q^cq$ (not displayed in Table~\ref{tab:mediators}), and can potentially destabilize the proton unless a protecting symmetry is introduced, see Ref.~\cite{2202.05275}.   Schematically, the matching of the pole residues to the concrete models in Table~\ref{tab:mediators}) take the following form:
    $[\cS_{I\,(a)}]_{\alpha\beta ij}=  [g^*_a]_{ij} [g^*_a]_{\alpha\beta}$,
    $[\cT_{I\,(b)}]_{\alpha\beta ij}= [g^*_b]_{i\alpha} [g^*_b]_{j\beta}$ and
    $[\cU_{I\,(c)}]_{\alpha\beta ij}=  [g^*_c]_{i\beta} [g^*_c]_{j\alpha}$.
Here $I\in\{V,S,T\}$ and $g^*_{a,b,c}$ denote generic couplings of the mediators to fermions of a given chirality and each index $a,b,c$ labels the possible mediator components contributing to the $s$, $t$ and $u$ channels, respectively.

%%%%%%%%%%%%%%%%%%%%%
\begin{table}[h!]
   \begin{center}
   \caption{ATLAS and CMS searches and their corresponding tail observables available in {\tt HighPT} package. These searches have been repurposed for generic New Physics entering Drell-Yan production.}
   \vspace{0.3cm}
    \renewcommand{\arraystretch}{1.}
    \begin{tabular}{|l | c| c| l|}
    \hline
        Process  & Experiment & Luminosity & Tail observable\\
        \hline
        $pp \to \tau\tau$ & ATLAS \cite{ATLAS:2020zms}  & $139\,\mathrm{fb}^{-1}$ & $m_T^{\rm tot}(\tau_h^1,\tau_h^2,\slashed{E}_T)$
        \\
        $pp \to \mu\mu$  & CMS \cite{CMS:2021ctt}  & $140\,\mathrm{fb}^{-1}$ & $m_{\mu\mu}$ 
        \\
        $pp \to ee$  & CMS \cite{CMS:2021ctt} & $137\,\mathrm{fb}^{-1}$ & $m_{ee}$ 
        \\\hline
        $pp \to \tau\nu$ & ATLAS \cite{ATLAS:2021bjk}  & $139\,\mathrm{fb}^{-1}$ & $m_T(\tau_h,\slashed{E}_T)$
        \\
        $pp \to \mu\nu$  & ATLAS \cite{ATLAS:2019lsy} & $139\,\mathrm{fb}^{-1}$  & $m_T(\mu,\slashed{E}_T)$ 
        \\
        $pp \to e\nu$  & ATLAS \cite{ATLAS:2019lsy} & $139\,\mathrm{fb}^{-1}$  & $m_T(e,\slashed{E}_T)$ 
        \\\hline
        $pp \to \tau \mu$  & CMS \cite{CMS:2022fsw} & $137.1\,\mathrm{fb}^{-1}$  & $m_{\tau_h\mu}^{\rm col}$ 
        \\
        $pp \to \tau e$  & CMS \cite{CMS:2022fsw} & $137.1\,\mathrm{fb}^{-1}$ & $m_{\tau_he}^{\rm col}$ 
        \\
        $pp \to \mu e$ & CMS \cite{CMS:2022fsw}  & $137.1\,\mathrm{fb}^{-1}$ & $m_{\mu e}$ \\
        \hline
    \end{tabular}
    \label{tab:lhc-searches}
    \end{center}
\end{table}
%%%%%%%%%%%%%%%%%%%%% 

%%%%%%%%%%%%%%%%%%%%%%%%%%%%%%%%%%%%%%%%%%%%%%%%%%%%%%%%%%%%%%%%%%%%%%%%%%
\section{Collider limits with {\tt HighPT}}
\label{sec:collider-limits}
%%%%%%%%%%%%%%%%%%%%%%%%%%%%%%%%%%%%%%%%%%%%%%%%%%%%%%%%%%%%%%%%%%%%%%%%%%

The high-energy regime of the dilepton invariant-mass or the monolepton transverse-mass are known to be very sensitive probes for a variety of New Physics models affecting semileptonic transitions \cite{Faroughy:2016osc,Greljo:2017vvb,Greljo:2018tzh,Marzocca:2020ueu,Angelescu:2020uug}. In the SM, the partonic cross-section scales as $\sim\!1/E^2$ at high-energies, leading to a smoothly falling tail for the kinematic distributions of any momentum-dependent observable. The presence of new particles coupling to quarks and leptons or new semileptonic interactions beyond the SM can modify the shapes of these tails substantially. The most obvious BSM effect is the appearance of a resonant feature on top of the smoothly falling SM background, i.e. a peak in the dilepton invariant mass spectrum, or an edge in the monolepton transverse mass spectrum. This indicates that a heavy colorless particle has been produced on-shell in the $s$-channel. On the other hand, non-resonant effects from contact interactions or on-shell leptoquarks exchanged in the $t/u$-channels can lead to more subtle non-localized features in the tails. Indeed, energy-enhanced interactions coming from non-renormalizable operators will modify the energy scaling of the distributions leading to an apparent violation of unitarity in the tails. The effects from leptoquarks exchanged in the $t/u$-channels lead to a similar behavior.

We introduce {\tt HighPT} \cite{Allwicher:2022mcg}, a {\tt Mathematica} package for the analysis of high-$p_T$ data in semileptonic processes at the LHC. The package is based on the form-factor framework introduced in Sec.~\ref{subsec:amp_dec}. With {\tt HighPT}, the user can compute Drell-Yan differential cross-sections and event yields in the tails  of experimental distributions in terms of the SMEFT Wilson coefficients at $\cO(1/\Lambda^4)$, including operators up to mass dimension-$8$, or in terms of specific New Physics models containing the bosonic tree-level mediators in Table \ref{tab:mediators}. The experimental searches implemented in the package are collected in Table \ref{tab:lhc-searches}. These correspond to data sets from the full Run-II ATLAS and CMS searches for heavy resonances in dilepton and monolepton production at the LHC. In the last columns we display the corresponding Drell-Yan tail observable measured in each search that is available in {\tt HighPT}. Specific details concerning the definition of the measured observables, selection cuts and any other inputs used in these experimental analyses are available in the respective ATLAS and CMS papers listed in second column of Table \ref{tab:lhc-searches}. More importantly, the user can also extract reliable limits including detector effects on the SMEFT and on mediator models. For each signal hypothesis, confidence intervals can be easily computed using Pearson’s $\chi^2$ test statistic ({\tt ChiSquareLHC}) which can then be minimized using the native {\tt Mathematica} routines for numerical minimization. 
%%%%%%%%%%%%%%%%%%%
\begin{figure}[h!]
    \centering
    \resizebox{1\textwidth}{!}{
    \begin{tabular}{c c c}
        \includegraphics[]{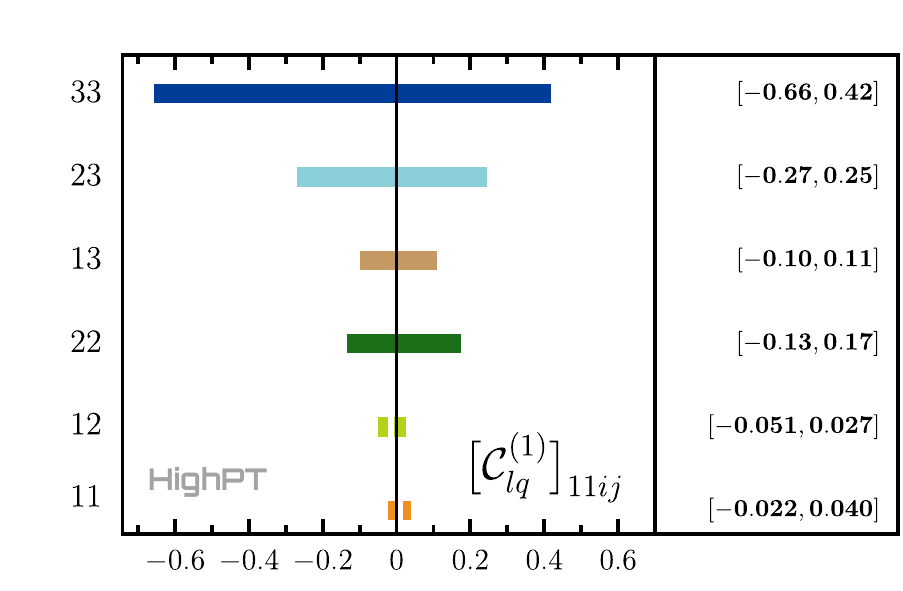} & \includegraphics[]{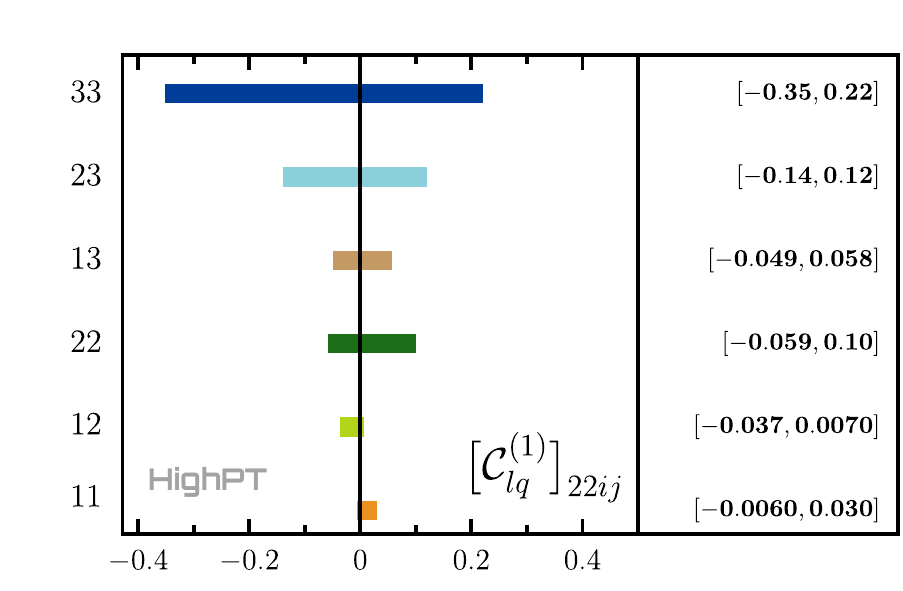} & \includegraphics[]{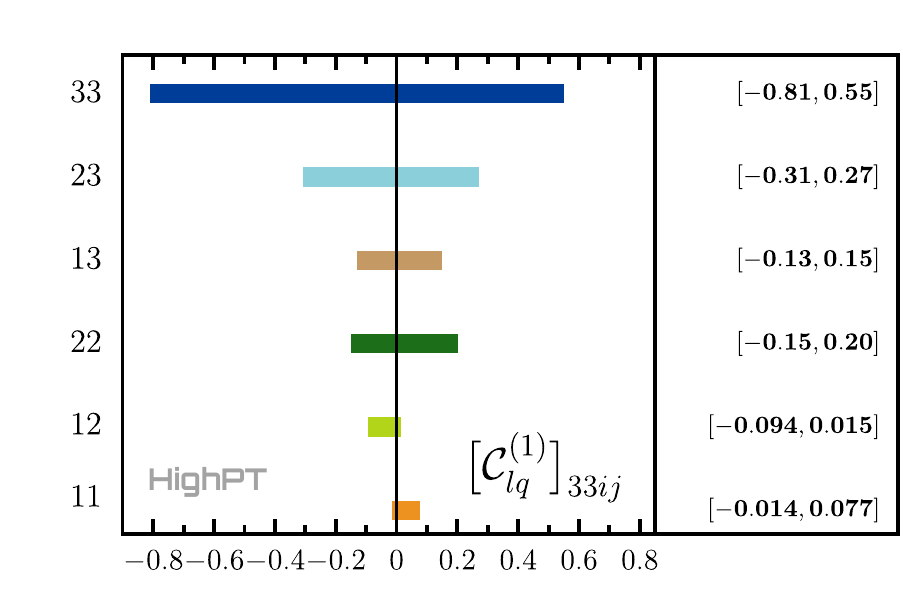} \\
        \includegraphics[]{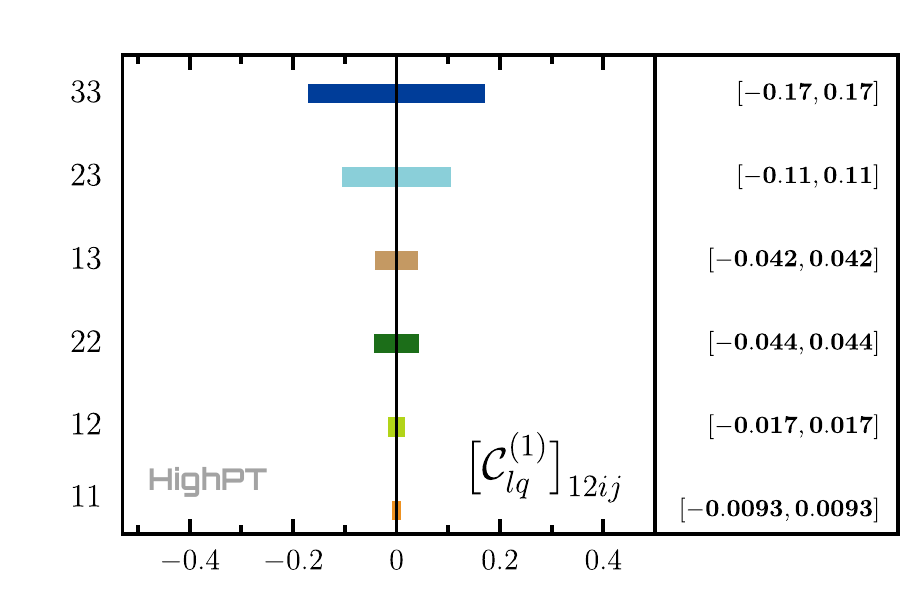} & \includegraphics[]{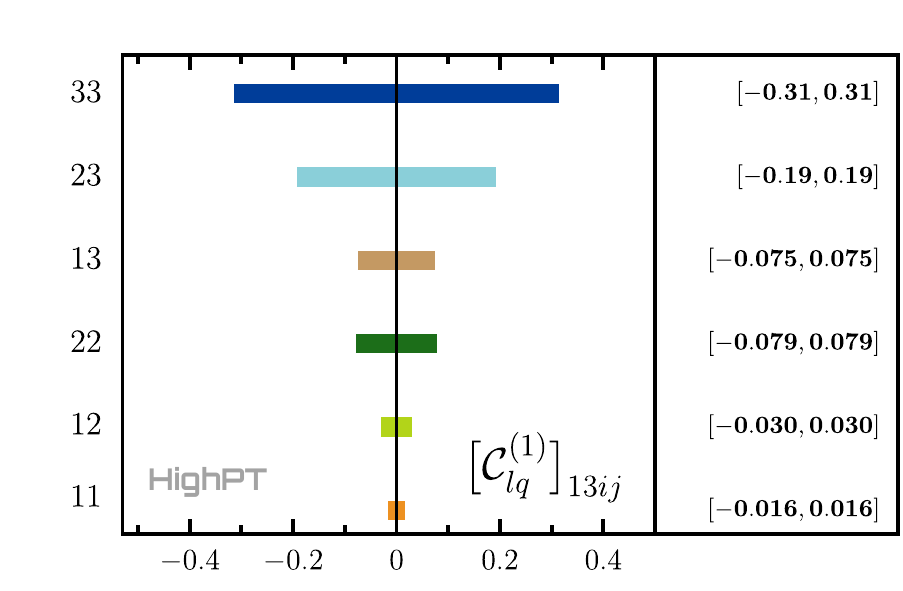} & \includegraphics[]{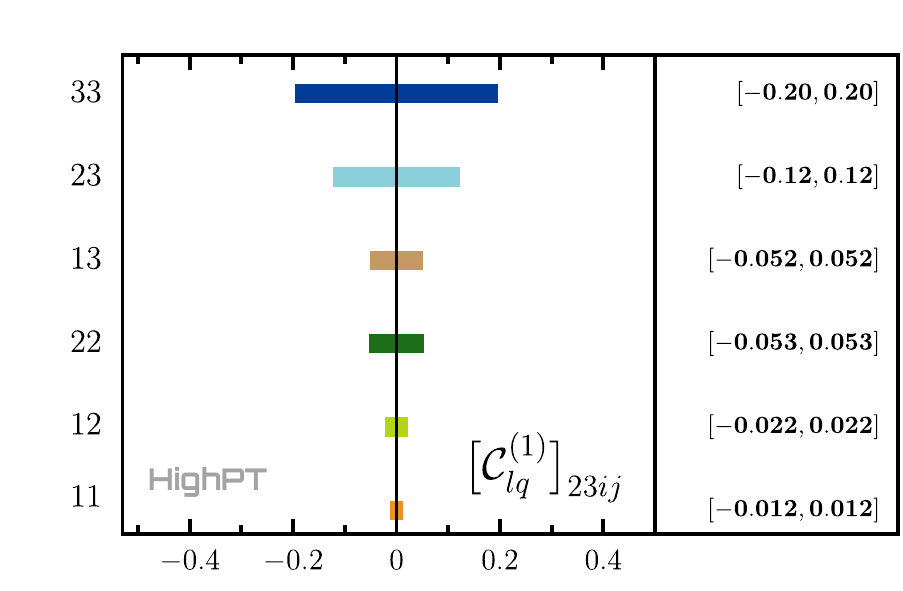} \\
        \includegraphics[]{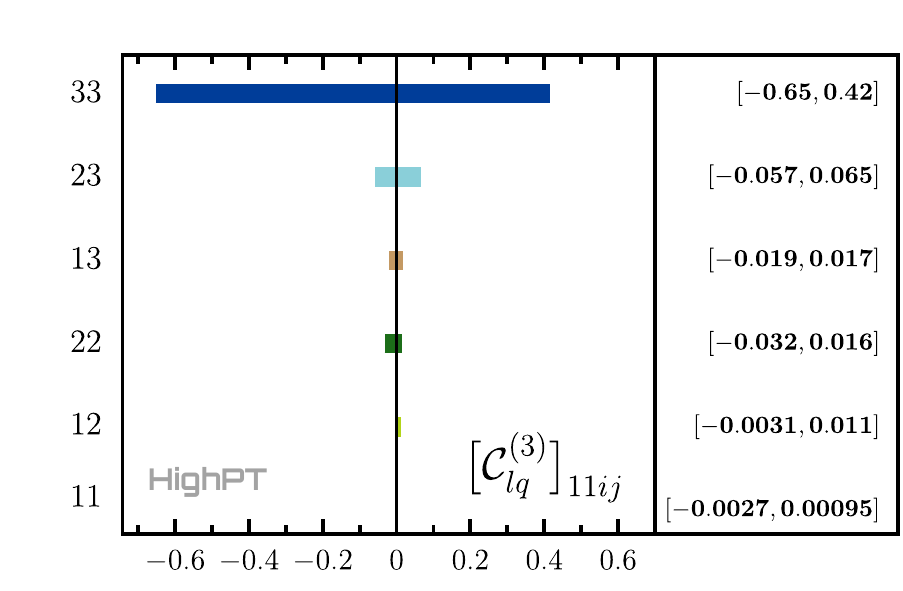} & \includegraphics[]{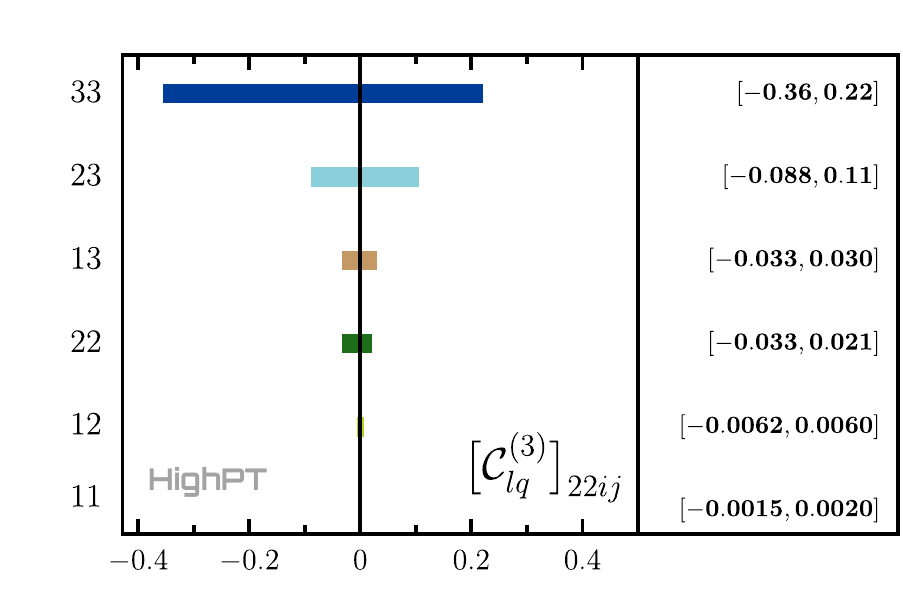} & \includegraphics[]{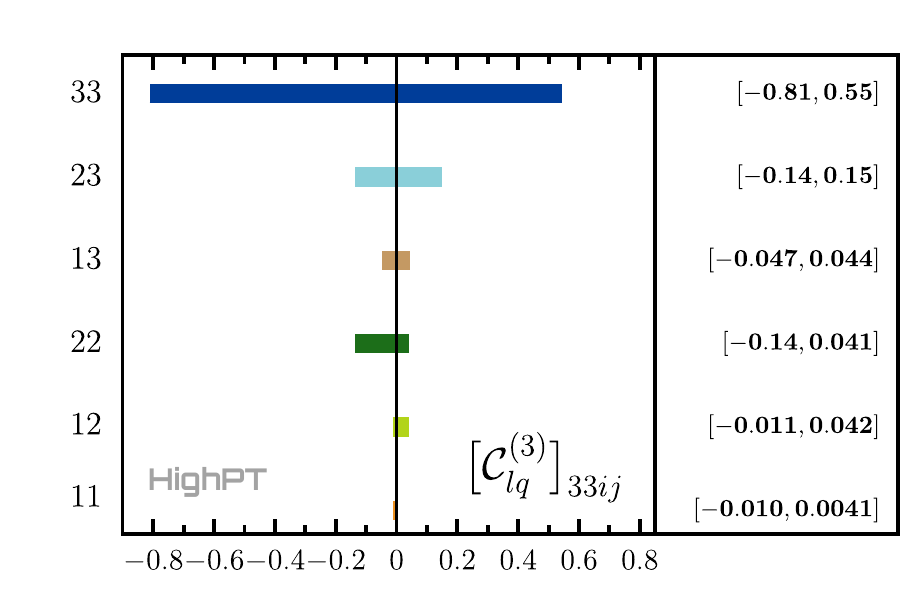} \\
        \includegraphics[]{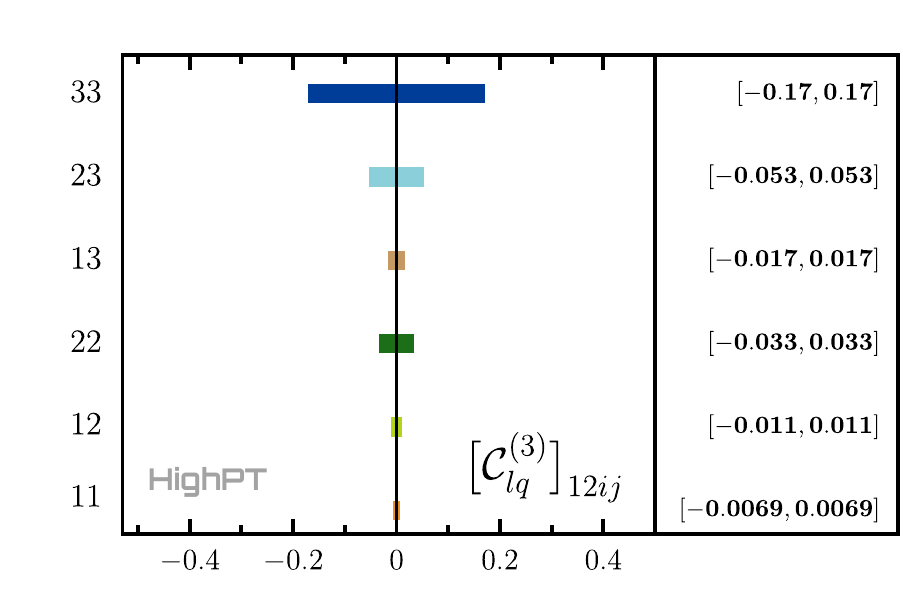} & \includegraphics[]{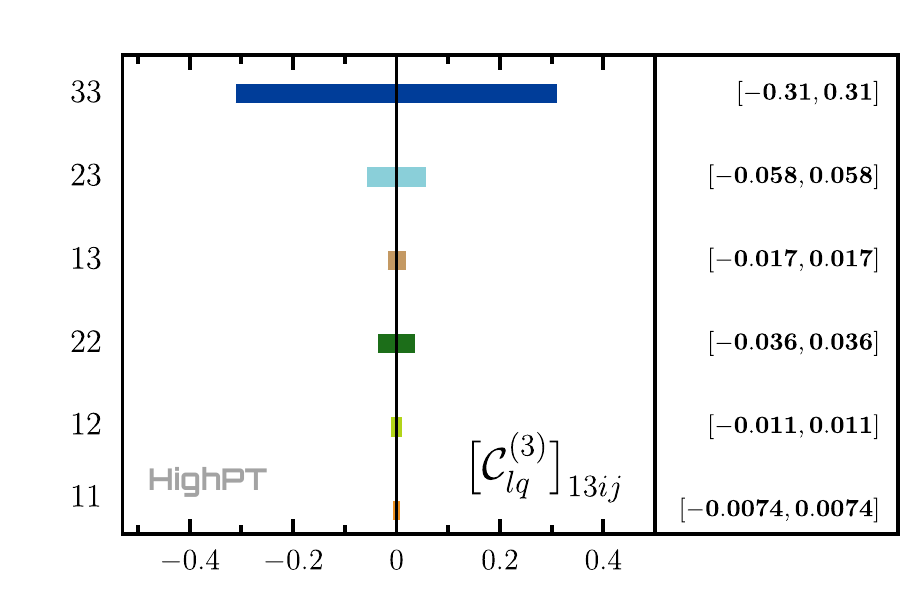} & \includegraphics[]{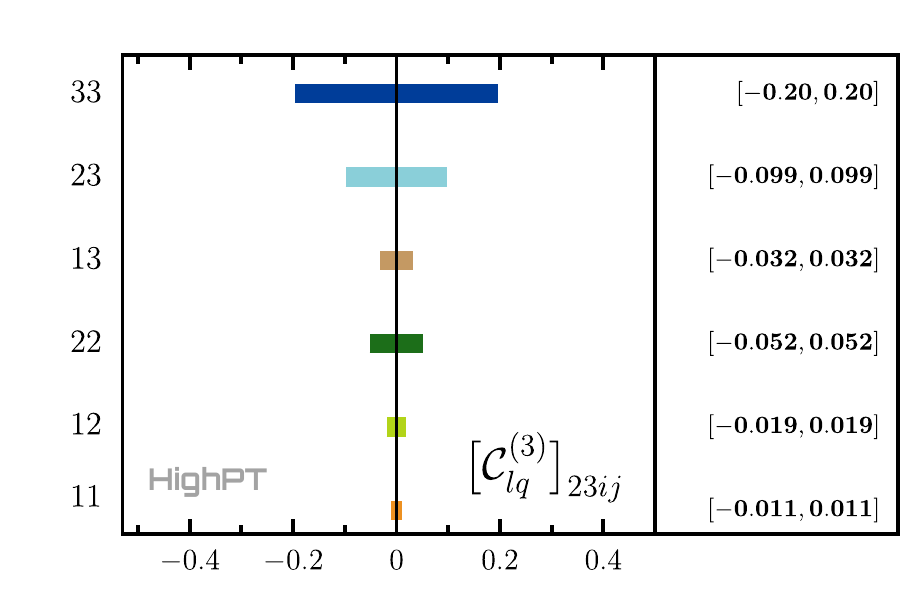}
    \end{tabular}
    }
    \caption{\sl\small LHC constraints at $95\%$ CL on the SMEFT Wilson $\cC_{lq}^{(1,3)}$ coefficients with different flavor indices, where a single coefficient is turned on at a time. Quark flavor indices are denoted by~$ij$ and are specified on the left-hand side of each plot. All coefficients are assumed to be real and contributions to the cross section up to and including $\cO(1/\Lambda^{4})$ are considered. The NP scale is chosen as~$\Lambda=1\,\mathrm{TeV}$.}
    \label{fig:single-WC-limits-lq}
\end{figure}
%%%%%%%%%%%%%%%%%%

%%%%%%%%%%%%%%%%%%%
\begin{figure}[b!]
    \centering
    \resizebox{1\textwidth}{!}{
    \begin{tabular}{c c c}
        \includegraphics[]{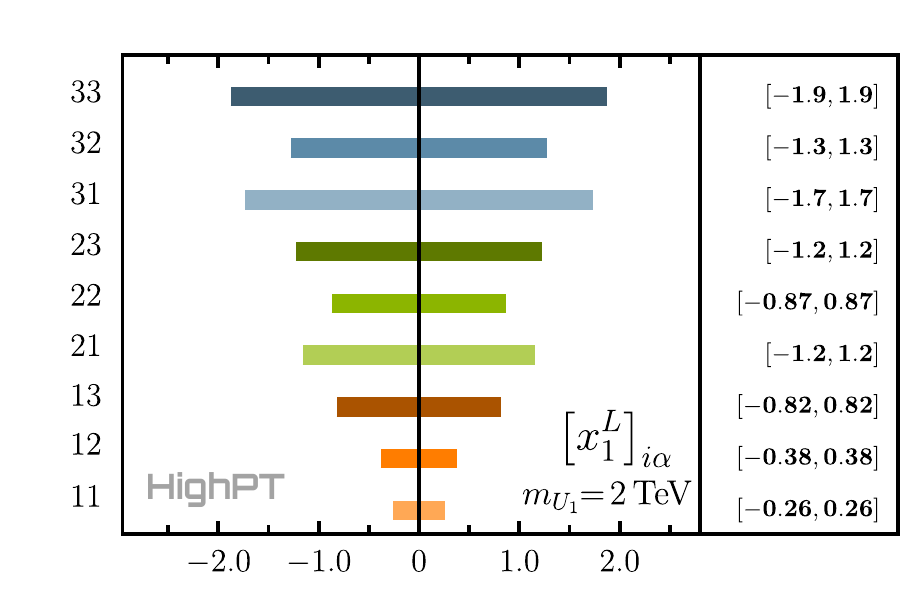} & \includegraphics[]{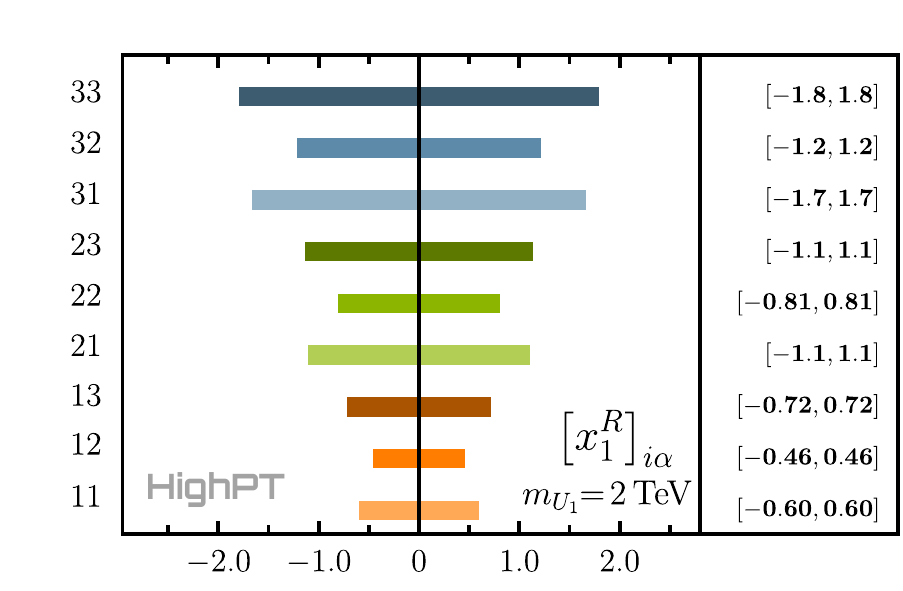} & \includegraphics[]{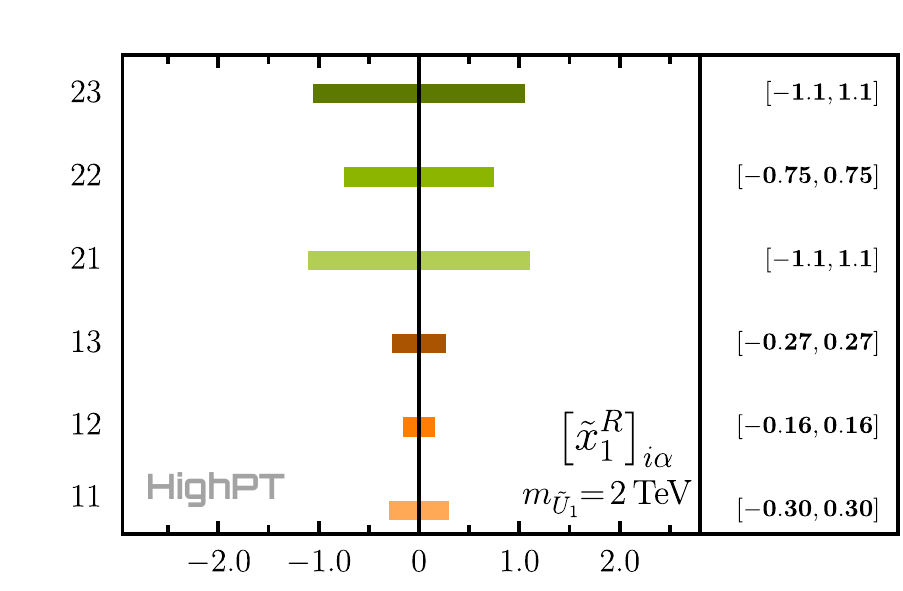}\\
        \includegraphics[]{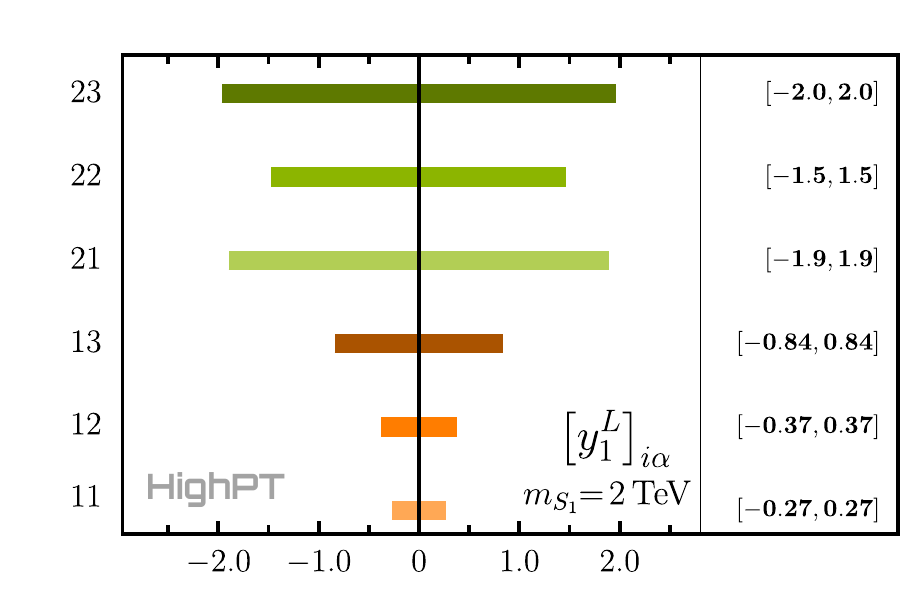} & \includegraphics[]{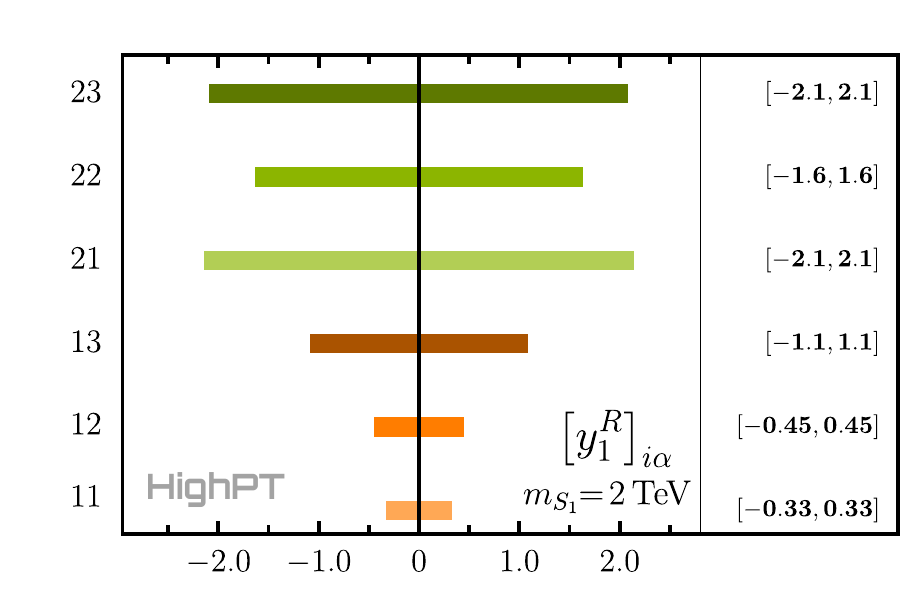} & \includegraphics[]{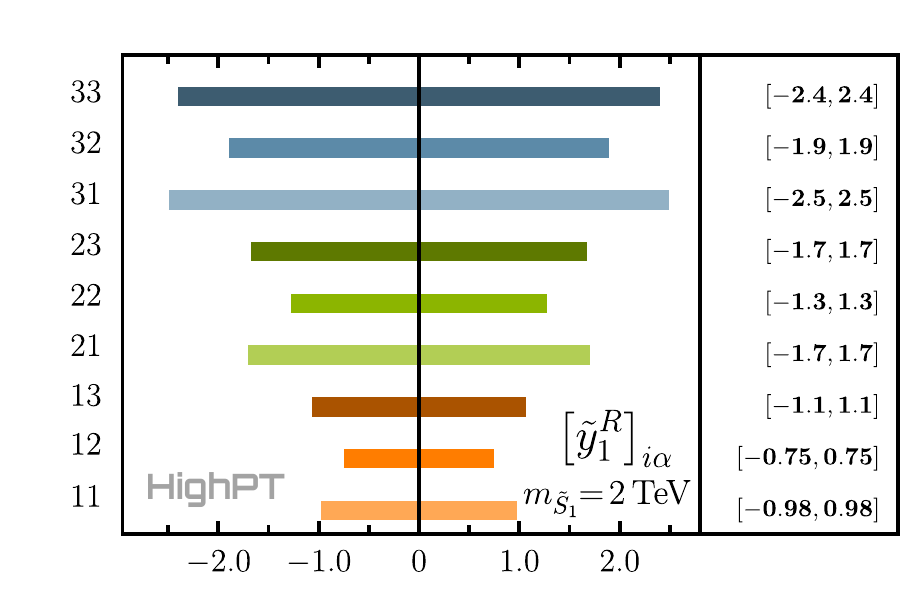}\\
        \includegraphics[]{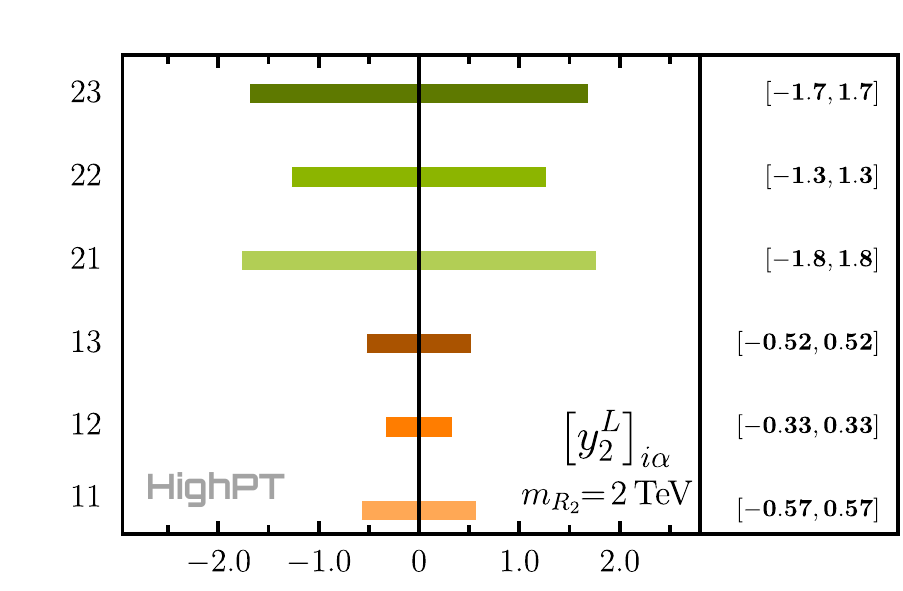} & \includegraphics[]{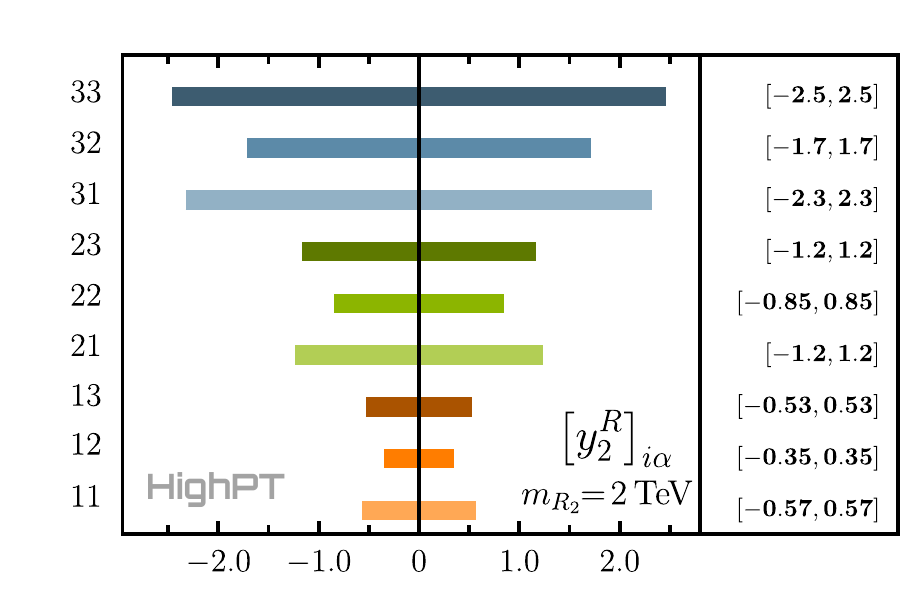} & \includegraphics[]{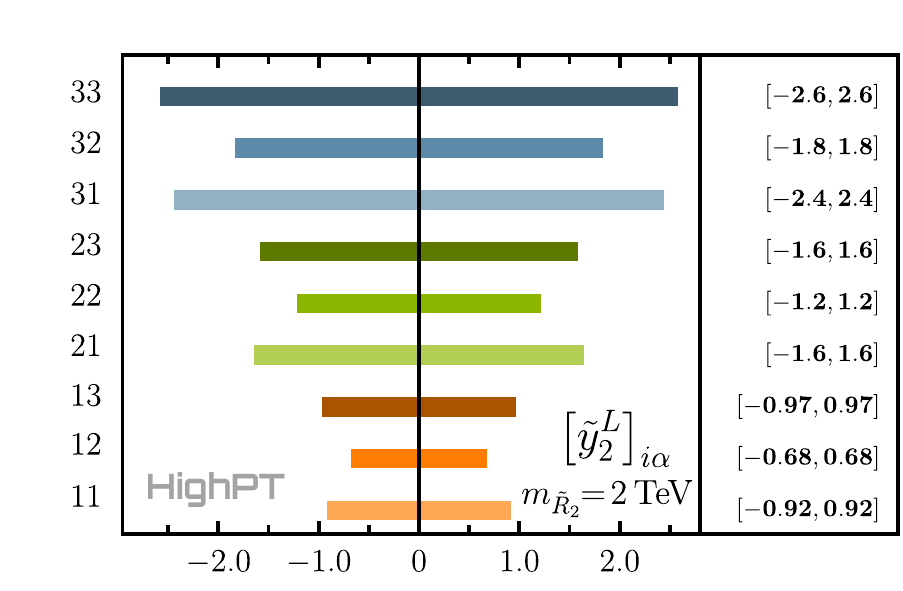}\\
    \end{tabular}
    }
     \caption{\sl\small LHC constraints at $95\%$ CL on the coupling constants of several leptoquark states, where a single coupling is turned on at a time. The masses of all leptoquarks are fixed to~$2\,\mathrm{TeV}$. The numbers on the left-hand side of each plot correspond to the respective flavor indices~$i\alpha$. See Tab.~\ref{tab:mediators} for the definition of the couplings.}\label{fig:LQ-couplings} 
\end{figure}
%%%%%%%%%%%%%%%%%%

\subsection{Single-operator limits on the dimension-$6$ SMEFT}
\label{ssec:smeft-limits}

In this Section we present upper bounds on the dimension-$6$ SMEFT operators using LHC Run-II data from the $pp\to\ell_\alpha^-\ell_\beta^+$ and $pp\to \ell_\alpha^{\pm}\nu_\beta$ Drell-Yan searches listed in Table~\ref{tab:lhc-searches}. Single-parameter limits are extracted for individual Wilson coefficients by assuming them to be real parameters and by setting all other coefficients to zero. For the sake of brevity, we only give results for the left-handed vector operators with the most general flavor structure:
\bea
\left[\cO_{lq}^{(1)}\right]_{\alpha\beta ij} &=& \left(\bar l_\alpha \gamma^\mu l_\beta\right)\left(\bar q_i \gamma_\mu q_j\right)\\
\left[\cO_{lq}^{(3)}\right]_{\alpha\beta ij} &=& \left(\bar l_\alpha \gamma^\mu\tau^I l_\beta\right)\left(\bar q_i \gamma_\mu\tau^I q_j\right)\,.
\eea
Limits for any other $d=6,8$ SMEFT operator that directly modify Drell-Yan production can be extracted with the {\tt HighPT} package. Our results are derived by keeping the $\cO(1/\Lambda^4)$ corrections from the dimension-$6$ squared pieces assuming flavor alignment in the down sector for the CKM matrix. The upper limits for the Wilson coefficients are presented in Fig.~\ref{fig:single-WC-limits-lq}. All limits are given at 95\% CL at a fixed reference scale of $\Lambda=1$\,TeV. Notice that for fixed leptonic flavors, as expected, we find that the most constrained coefficients are the ones involving valence quarks, but useful constraints are also obtained for operators involving the heavier $s$-, $c$- and $b$-quarks despite the PDF suppression. Overall, the upper limits for $\cO_{lq}^{(1,3)}$ for different $(i,j)$ indices follow approximately the expected hierarchies between the parton-parton luminosity functions. For fixed quark flavor-indices, we find comparable constraints between the $e$ and $\mu$ channels, with much weaker constraints for $\tau$'s.

\subsection{Concrete models: Leptoquarks}
\label{ssec:smeft-limits}

We now provide limits on the couplings of leptoquark states. We focus on three examples: the $u$-channel scalar singlets $S_1$ and $\widetilde{S}_1$, the $t$-channel scalar doublets $R_2$ and $\widetilde{R}_2$, and $t$-channel vector singlets $U_1$ and $\widetilde{U}_1$. We consider a single coupling at a time (dropping RH neutrinos as final states) and take for each leptoquark a benchmark mass of $2$~TeV. Our results are collected in Fig.~\ref{fig:LQ-couplings} for the leptoquark couplings that contribute to dilepton and monolepton tails, where we show the $95\%$ confidence intervals for each individual coupling. For fixed quark flavors, these results follow the same pattern of the SMEFT results presented above, with the strongest bounds corresponding to the lightest quarks with the larger PDFs.

\section{Summary}
In this talk we have presented a high-$p_T$ analysis of semileptonic New Physics entering charged and neutral Drell-Yan production at the LHC.
Starting with a general description of the scattering amplitude in terms of form-factors, we introduced a useful parametrization and discussed how this framework can be matched to specific BSM models such as the SMEFT truncated at $\cO(1/\Lambda^{4})$ (including dimension-$8$ effects), as well as explicit UV models with new (tree-level) bosonic mediators that can be resolved by the collider energies. Using data from run-II LHC searches in $pp\to \ell\ell^{(\prime)}$ and $pp\to \ell\nu$, we have presented the most stringent high-$p_T$ limits on SMEFT Wilson coefficients, as well as leptoquark models, with arbitrary flavor structures. These single-parameter limits were extracted with {\tt HighPT}, a dedicated Mathematica package for high-$p_T$ collider analyses for generic BSM physics entering semileptonic transitions.     

\section*{Acknowledgments}
This work has received funding from the European Research Council (ERC) under the European Union’s Horizon 2020 research and innovation programme under grant agreement 833280 (FLAY), and by the Swiss National Science Foundation (SNF) under contract 200021-175940

\end{document}